\documentclass[9pt,twocolumn,twoside]{pnas-new}

\templatetype{pnasresearcharticle} 

\usepackage{graphicx}
\usepackage{natbib}
\usepackage[section] {placeins}
\usepackage{subfigure}
\usepackage{float}
\usepackage{color}
\usepackage{amsmath}
\usepackage{comment}

\usepackage[scaled]{helvet}

\usepackage[T1]{fontenc}
\usepackage{color}

\usepackage{xcolor} 
\definecolor{bg-green}{rgb}{0.8588,0.9333,0.8666}
\definecolor{Q-color}{rgb}{0.5,0.1,0.5}
\newcommand{\yq}{}

\def\apj{Astrophysical Journal}
\def\araa{Annual Review of Astronomy and Astrophysics}             
\def\apjl{Astrophysical Journal Letters}                
\def\apjs{Astrophysical Journal Supplement}               
\def\aap{Astronomy and Astrophysics}                
\def\aapr{Astronomy and Astrophysics Reviews}          
\def\mnras{{Monthly Notices of the Royal Astronomical Society}}             
\def\nat{{Nature}}              

\def\msun{{\rm\,M_\odot}}

\newcommand{\kms}{\, {\rm km\, s}^{-1}}

\newcommand{\mpc}{\, {\rm Mpc}}

\def\h2{${\rm\,H_2}$}

\newcommand{\erg}{\,\mathrm{erg}}

\newcommand{\K}{\,\mathrm{K}}

\newcommand{\alf}{Alfv$\acute{\text{e}}$n\ } 

\makeatletter 

\def\Mpc{{\rm\,Mpc}}
\def\mpc{{\rm\,Mpc}}

\def\msun{{\rm\,M_\odot}}

\def\vol#1  {{{#1}{\rm,}\ }}

\newcount\refno
\newcount\rfno
\def\eq{$^{\the\refno\ }$\advance\refno by 1}
\def\ad{\advance\rfno by 1}

\def\clock{\count0=\time \divide\count0 by 60
     \count1=\count0 \multiply\count1 by -60 \advance\count1 by \time
     \number\count0:\ifnum\count1<10{0\number\count1}\else\number\count1\fi}

\def\myputfigure#1#2#3#4#5%
{\hskip0.03\textwidth\vskip#5pt
\makebox[0pt]{\hskip#2in
\includegraphics[width=#3\textwidth]{#1}}\vskip#4pt\hfill}

\begin{document}

\title{Reduced Gas Accretion onto Galaxies due to Effects of External Giant Radio Lobes}

\author[a,b]{Yu Qiu}
\author[a,b,*]{Renyue Cen}

\affil[a]{Center for Cosmology and Computational Astrophysics, Institute for Advanced Study in Physics, Zhejiang University, Hangzhou 310027, China}
\affil[b]{Institute of Astronomy, School of Physics, Zhejiang University, Hangzhou 310027, China}

\leadauthor{Qiu \& Cen}

\significancestatement{
Powerful radio jets released by accreting supermassive black holes are a significant source of magnetic energy in the Universe.  By cosmic noon (redshift 2 to 3), the jet-inflated radio lobes extend well beyond their host galaxies and inject a substantial amount of magnetic energy into a significant portion of the intergalactic medium. Subsequently, gas accretion onto nearby galaxies is hindered through magnetic pressure, which provides a global negative feedback mechanism. We use cosmological magneto-hydrodynamic simulations to quantify this effect and find significant reduction of gas accretion onto halo of masses ranging from $10^{11}$ to $10^{14}\msun$. This external AGN feedback process has significant bearings on galaxy formation and evolution. Accurate modeling of galaxy formation demands the incorporation of this new, external AGN feedback process. 
}

\authorcontributions{Author contributions: R.~C. formulated the research and designed the scientific goals and metrics of the simulations.
Y.~Q. generated, performed and analyzed the simulations.
Y.~Q. and R.~C. wrote the paper.}
\authordeclaration{The authors declare no competing interest.}
\correspondingauthor{\textsuperscript{*}E-mail: renyuecen@zju.edu.cn}


\begin{abstract}
Suppression effects of giant radio lobes from supermassive black holes on gas accretion onto galaxies in the surrounding regions are quantified using cosmological magneto-hydrodynamic simulations. With an appropriate amount of radio jet energy injected into the intergalactic medium following the formation peak of supermassive black holes at redshift two, we find that galaxies in the greater neighborhood of the jet-launching massive galaxies subsequently experience a significant reduction in the amount of accreted gas. The distribution of the resulting magnetic field in the intergalactic medium is highly inhomogeneous, due to the highly biased nature of the most massive supermassive black holes. In regions with magnetic field strength $B>10^{-2}\mu$G, the baryon fraction is on average reduced by 17\%, 14\%, and 12\%, respectively, for halos of mass in the range of $[10^{11}-10^{12})\msun$, $[10^{12}-10^{13})\msun$, and $[10^{13}-10^{14})\msun$. A proper inclusion of this new, external, global, preventive feedback mechanism from AGN in the next generation of cosmological simulation may be necessary.
\end{abstract}

\dates{This manuscript was compiled on \today}
\doi{\url{www.pnas.org/cgi/doi/10.1073/pnas.XXXXXXXXXX}}

\maketitle
\thispagestyle{firststyle}
\ifthenelse{\boolean{shortarticle}}{\ifthenelse{\boolean{singlecolumn}}{\abscontentformatted}{\abscontent}}{}

\firstpage[3]{6}


\dropcap{M}agnetic fields are ubiquitous in and around galaxies \citep{Grasso2001, Govoni2004, Beck2015}, as evidenced by observations of the radio synchrotron emission 
\citep{Feretti2012, Han2017}. 
Primordial magnetic fields generated in cosmic inflation and phase transitions \citep{Ratra1992, Durrer2013, Subramanian2016}
may be amplified by homogeneous isotropic Kolmogorov turbulence associated with gravitational structure formation of galaxies \citep{1997Kulsrud} or dynamo mechanism within galaxies \citep{Brandenburg2005}.


Most cosmological simulations that involve magnetic fields either adopt a primordial ``seed'' field \citep{Vazza2014, 2018Springel, Katz2021}, or focus on the generation of magnetic fields from galactic outflows driven by starbursts \citep{Bertone2006, Donnert2009}. Recently, Cen 2024 \citep{Cen2024} showed that the intergalactic medium (IGM) endowed with a substantial amount of magnetic energy originating from giant radio lobes, i.e., Fanaroff-Riley Class II (FR-II) jets, may be hindered from being accreted onto galaxies.
The hindering effect is analogous to thermal pressure, except that the magnetic energy trapped in the gas is dissipationless and long-lived.
\yq{A similar non-thermal pressure support can be provided by cosmic-ray protons \citep[e.g.,][]{Ruszkowski2023}, which may help remove baryons from galaxies \citep[e.g.,][]{Quataert2025}.}
The expectation is that, when magnetic pressure in the accreting gas becomes comparable to the \yq{thermal pressure (or equivalently, in a virialized gas, the gravitational energy density)}, when approaching the virial radius of a host halo, gas will be significantly hindered from accreting onto the galaxy. 

In contrast with traditonal ``internal'' feedback models implemented in a variety of cosmological simulations \citep[e.g.,][]{2014Vogelsberger,2014Genel,2015Schaye,2015Crain,2015Schaller,2015Trayford,2016McAlpine, 2018Pillepich,2018Springel,2018Nelson, 2019Dave,2022Sorini}, this radio lobe AGN feedback mechanism is external,  global, and preventive in nature.
Because FR II jets transport magnetic energy into the low-density IGM, far away from originating host galaxies, 
they not only energetically contain a substantial fraction of the supermassive black hole rest mass energy,
but also provide the most economical way to maximize the generation of entropy in the IGM per unit energy deposited.
In this article, we use cosmological magneto-hydrodynamic simulations to quantify the impact of powerful radio jets on the accretion of baryons onto galaxies, as well as its spatial and environmental dependence.

\section*{Magneto-hydrodynamic simulations in a cosmological volume}

The simulations are performed using the cosmological MHD code Enzo \citep{Bryan2014}, utilizing the ideal magneto-hydrodynamics module with constrained transport \citep{Li2008, Collins2010}, which preserves the divergence of the magnetic field.
Starting at redshift $z=100$, the initial conditions are generated using the code MUSIC \citep{Hahn2011}. 
We adopt cosmological parameters from Planck 2018 results \citep{Aghanim2020}, i.e., a flat universe with Hubble constant $H_0\equiv100\,h\kms\mpc^{-1}=67.4\kms\mpc^{-1}$, total non-relativistic matter density parameter $\Omega_m=0.3138$, baryon density parameter $\Omega_b=0.0493$, power spectrum amplitude $\sigma_8=0.811$, and spectral index $n_s=0.965$. 

The simulation is periodic in a box of size $20\,h^{-1}\,\mathrm{cMpc}$ on a uniform $1024^3$ grid. 
The largest cluster at $z=0$ in the simulation box has a halo mass $M_h\approx10^{14}\msun$.
The resolution is designed to resolve the virial radius of $M_h=10^{11}\msun$ halos with about five cells.
\yq{This allows us to probe the baryon accretion at the virial radii, without significantly increasing the computational cost to resolve the interior of halos. Increasing the resolution potentially enables the formation of more detailed magnetic and turbulent velocity structures on small scales, which can strengthen the non-thermal pressure support. Therefore, we anticipate the relatively low resolution simulation will provide a lower limit on the baryon reduction effect. }
No primordial magnetic field is used in the simulations. 
A tabulated cooling table for the primordial gas is generated using the GRACKLE library \citep{Smith2017a}. 
\yq{Because the typical cooling timescale at the virial radii of halos is longer than the Hubble time, cooling associated with metal species does not significantly affect the baryon accretion at halo boundaries. Therefore, we do not model metal enrichment and associated cooling in the simulations.}
A baryon density floor of $10^{-2}\,\overline{\rho}_m$, where $\overline{\rho}_m$ is the mean matter density at a given redshift, and a temperature floor of $5000\K$ are introduced to avoid significant slowdown of the simulation timestep. Compared to simulations with the floors removed, as well as with a six-species (i.e., ionized and neutral hydrogen and helium) non-equilibrium cooling module in GRACKLE, we have verified in low-resolution simulations that the relevant properties of the resulting halos are not affected.

\section*{Implementation of giant radio lobe feedback}
As laid out in \citep{Cen2024}, the external feedback process is implemented as jet-inflated magnetic bubbles in the simulations. The cosmological volume is initially evolved magneto-hydrodynamically with a token (i.e., zero) magnetic field
to $z=2$. A total of $N_h=48$ $M_h>10^{12}\msun$ halos are then located using the ROCKSTAR halo finder \citep{Behroozi2013}, with a total mass of $\Sigma_i{M_{h,i}}\sim 10^{14}\msun$. The halo mass threshold ($M_h>10^{12}\msun$) is chosen so as to identify potential hosts of supermassive black holes (SMBHs) with masses $M_{\rm BH}>10^8\msun$, whose feedback energy is substantial. The total feedback energy $E_{\rm fb}$ is then calculated based on the formula:
\begin{equation}
\begin{split}
E_{\rm fb} &= 2\,N_h\,E_{\rm bub}\\
&= \eta_*\,\eta_{\rm BH}\,\eta_{\rm R}\,\eta_{\rm J}\,\frac{\Omega_b}{\Omega_m}\,\Sigma_i{M_{h,i}}\,c^2\\
&= 1.6\times 10^{60}\,\left({2\,N_h}\right)\,\left(\frac{\eta_*}{0.2}\right)\,\left(\frac{\eta_{\rm BH}}{0.002}\right)\,\left(\frac{\eta_{\rm R}}{0.2}\right)\,\left(\frac{\eta_{\rm J}}{0.07}\right)\,\\
& \left[\frac{\Omega_b/\Omega_m}{0.157}\right]\,\left(\frac{\Sigma_i{M_{h,i}}}{10^{14}\msun}\right)\,{\rm erg},\\
\end{split}
\label{eqn:energy}
\end{equation}
where $E_{\rm bub}$ is the total energy of each magnetic bubble (assuming each SMBH releases a pair of bubbles), various parameters are, respectively, 
stellar to baryon mass ratio ($\eta_{*}$), 
SMBH to stellar mass ratio ($\eta_{\rm BH}$), 
fraction of all relevant mass galaxies classified as radio loud galaxies ($\eta_{\rm R}$)
and the mean jet power in units of accretion power onto the SMBH in radio-loud galaxies ($\eta_{\rm J}$). 
Fiducial values of the parameters in the equation are chosen based on observations \citep[see a discussion in][]{Cen2024} and used here. 
\yq{In general, radio observations of FR II jets report total energy $E_{\rm bub}\sim10^{60}\erg$ per lobe \citep{ODea2009} under minimum energy magnetic field assumption \citep{Miley1980}, consistent with our energy modeling.} 
While the simulation domain is initialized with ${\bf B}=0$, in order to maintain $\nabla\cdot {\bf B}=0$ after injecting feedback energy, the magnetic field in the bubble is implemented as the curl of a vector potential field, i.e., ${\bf B}_{\rm bub}=\nabla\times {\bf A}_{\rm bub}$. 
The initial distribution of the magnetic field in a bubble follows the Kolmogorov power spectrum. 
\yq{A spherical error function mask is applied to ${\bf A}_{\rm bub}$, so that after taking the curl, a Gaussian smoothing is applied to ${\bf B}_{\rm bub}$ to circumvent abrupt variation in magnetic field strength at the bubble surface. }
Based on isolated bubble expansion from an initial small radius to 1 Mpc radius, $\eta_B\approx3\%$ of the initially injected magnetic energy remains in magnetic form at $r=1$\,Mpc, with the rest having been converted into kinetic or thermal energy, i.e.,
$E_{{\rm mag}}(r=1\,{\rm Mpc}) = \eta_B E_{\rm bub}$.
\yq{The adopted radius, 1\,Mpc, is on the high end of observed radio lobes \citep[e.g.,][]{Hardcastle2000}, which represents the later, mature stage of radio lobe evolution \citep[see, e.g.,][for the early stage inflation of the lobe from collimated jets and subsequent turbulent evolution]{Tchekhovskoy2016, Vazza2021}. For lobes starting with a smaller radius, e.g., $100\,{\rm pkpc}$, it would take $\sim 1$\,Gyr, given the lobe energy and ambient gas density, to grow to $1\,{\rm Mpc}$, which closely follows the Sedov-Taylor solution based on simulations we have performed. Therefore, the results would not change significantly if we insert $r\sim100$\,kpc bubbles at $z=3$, as the bubbles will expand to $1\,{\rm Mpc}$ by $z=2$. In order to focus the computational resources on the large-scale impact, the 96 magnetic bubbles with radius $r=1\,{\rm Mpc}$} are inserted in the simulation around their respective host halos at $z=2$.
Each bubble pair is oriented randomly on opposite sides of the host halo, offset $1\,{\rm Mpc}$ from the halo center.
\yq{The direction of deposition affects which part of the IGM the radio lobe may interact with contemporaneously. 
In practice, in the proto-cluster environment, many pairs of jets will be deposited. Still, it may be that the volume filling fraction of all lobes may be less than 100\%, even in proto-cluster regions, which will cause inhomogeneous feedback effects. 
The two scientific goals of this study are to demonstrate (1) the significance of this feedback process and (2) its 
spatially inhomogeneous nature. We note that this study does not address the directional nature of the lobes with respect to the jet launching host galaxy and the internal feedback effect.}
The bubble inserted simulation box is restarted at $z=2$ running to $z=0$.

\section*{Suppression of baryon fraction in halos by giant radio lobes}

\begin{figure*}
\centering
\includegraphics[width=1.0\linewidth]{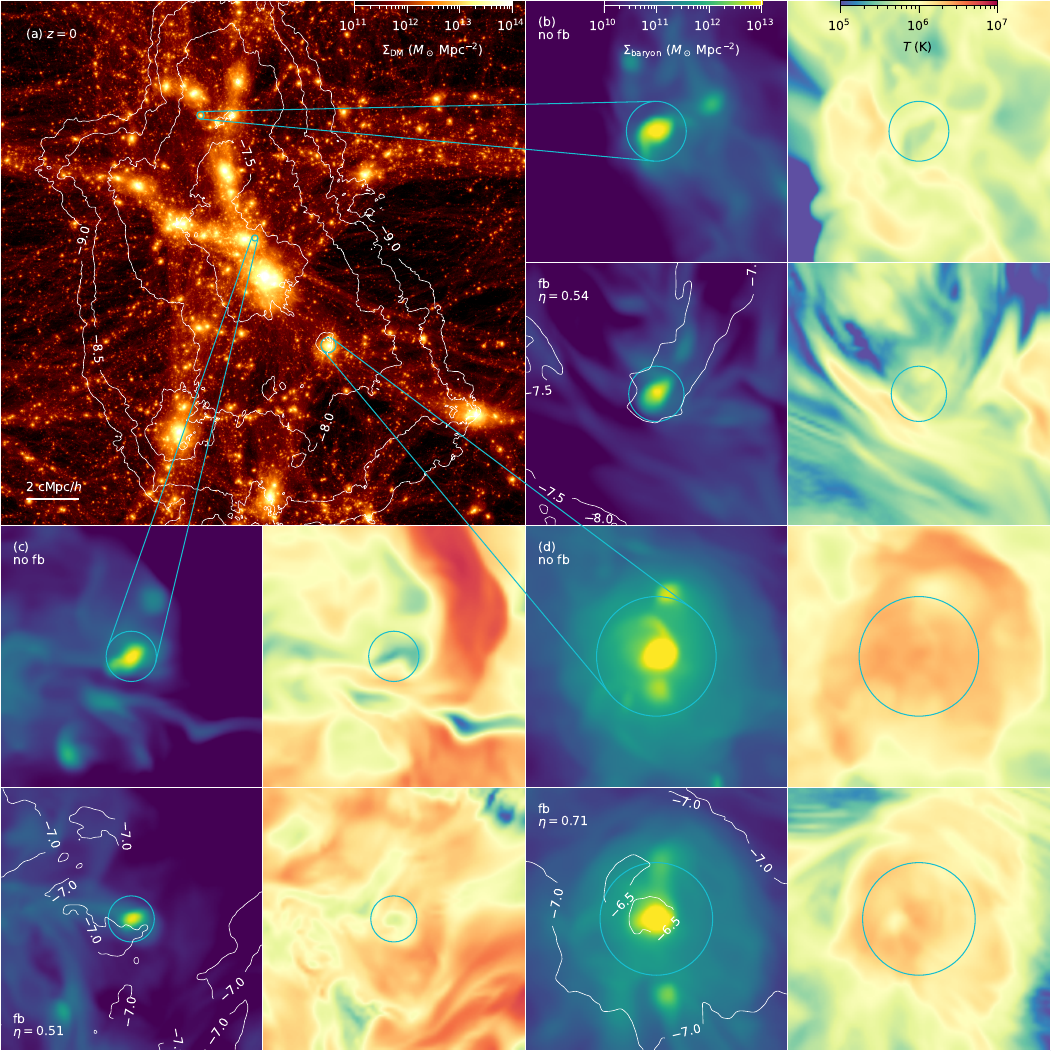}
\caption{
Top-left panel (a) shows the projected distribution of the dark matter in the 20 cMpc/h simulation box with magnetic bubbles at $z=0$. Volume-weighted projected magnetic field magnitude is overlayed as white contours ($\log{|B|}$ in units of Gauss); the four visible contour levels 
are (-9.0, -8.5, -8.0, -7.5) in panel (a). Additional contours of higher magnitudes are visible in zoomed panels (b,c,d).
The three four-panel sets zoom into three separate halos that show significant baryonic mass reduction due to magnetic bubbles, with $\eta=0.54$ (panel b), $0.51$ (panel c), and $0.71$ (panel d), respectively.
In each four-panel set, the top-row of two panels show the gas density (left panel) and temperature (right panel) from the simulation without magnetic bubbles, indicated as "no fb". The bottom-row panels show the corresponding variables from the simulation with magnetic bubbles, indicated as "fb". 
The cyan circles indicate the virial radius of each halo. The widths of panels (b,c,d) are 2\,Mpc, and the projection depths are three times their respective virial radius centered at the halo.
}
\label{fig:box}
\end{figure*}

\begin{figure*}
\centering
\includegraphics[scale=1.0]{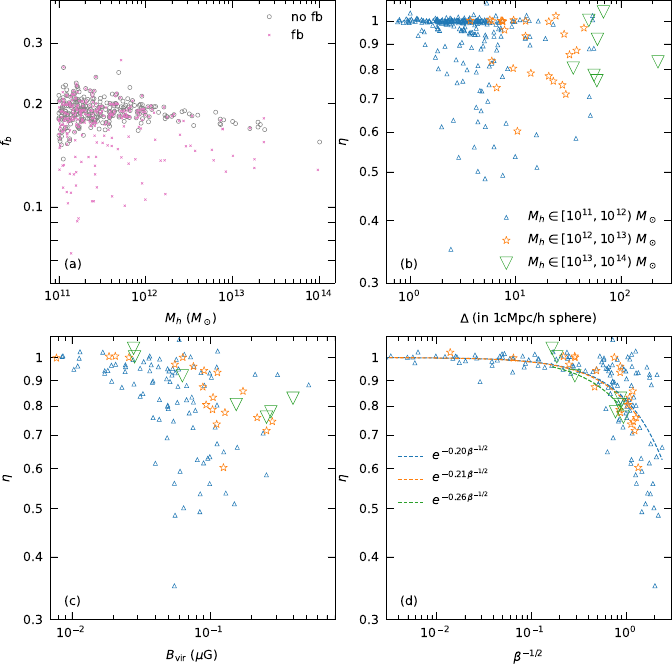}
\caption{Panel (a) shows the comparison of baryon fraction $f_b$ as a function of halo mass $M_h$, in simulations with or without feedback. \yq{Halos not affected by feedback have $f_b$ higher than cosmic mean due to radiative cooling.} The rest of the panels show the normalized baryon fraction $\eta$ as a function of (b) mean density $\Delta$, (c) volume-weighted magnetic field magnitude between $(1,1.2)\times r_{\rm vir}$, and (d) $\beta^{-1/2}$ at the virial radii. Exponential fits in the form of $e^{-\alpha \beta^{-1/2}}$ are shown in the panel (d), which indicate that baryons are prevented from entering the halos due to magnetic pressure force.}
\label{fig:etascatter}
\end{figure*}

Panel (a) of Fig.~\ref{fig:box} shows the distribution of dark matter at $z=0$ in the simulation with giant radio lobe feedback. The magnetic field is displayed as white contours in the plot, which are seen to approximately center on the most massive $M_h\sim10^{14}\msun$ halo in the simulation at the center of the display, 
due physically to the fact that jet launching massive galaxies are highly biased.
To remove ambiguity in defining the virial radius of a halo,
only central halos are analyzed here (i.e., excluding subhalos and halos experiencing major mergers). 
Three illustrative halos in the feedback run (labeled ``{\rm fb}'') are shown in panels (b,c,d) of Fig.~\ref{fig:box}, spanning a wide range in halo mass and distance to the central halo in the simulation.
Baryon density and temperature projections are shown for each halo, and compared with the scenario without feedback (labeled ``no fb''). 
\yq{The baryon fractions for all halos studied in this work, $f_b$, are shown in Fig.~\ref{fig:etascatter} as a function of halo mass. Due to the absence of ``internal'' feedback mechanisms in our study, in regions not affected by the radio lobes, the baryon fraction may be above the cosmic mean due to excessive radiative cooling \citep[also seen in cosmological simulations with radiative cooling, e.g.,][]{Kravtsov2005}.
We have verified by running simulations without radiative cooling that this above-the-cosmic-mean baryon fraction is due to radiative cooling.
To isolate the baryon reduction effect from giant radio lobes,} a normalized baryon fraction parameter, $\eta$, defined as the ratio of baryon
fractions of the halo in the simulation with feedback to that of the corresponding halo in the simulation without feedback, is shown in white for each halo. 
In the three cases shown in Fig.~\ref{fig:box}, the baryon content is significantly reduced by 29\%-49\%, with central magnetic field strength approaching $0.1-1\mu$G. 
It is evident that significant external magnetic fields can substantially reduce gas accretion onto galaxies. 

\begin{figure*}
\centering
\includegraphics[scale=1.0]{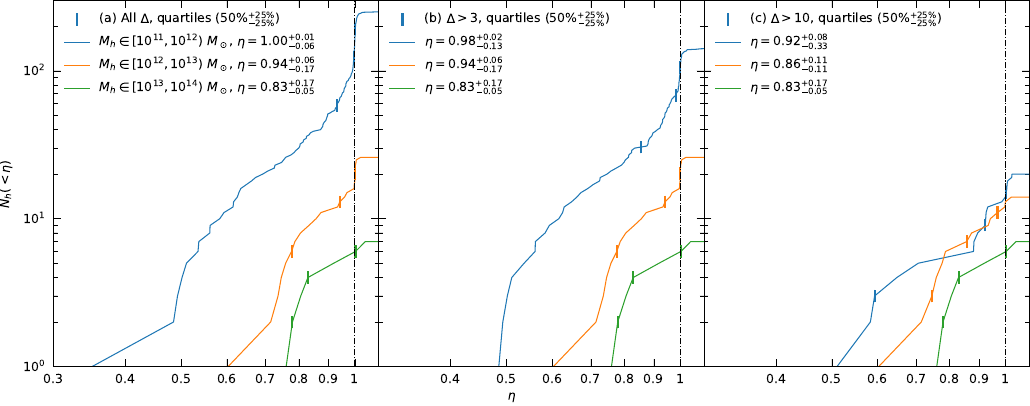}
\caption{Cumulative distribution of $\eta$ in different $M_h$ ranges in the simulation with giant radio lobe feedback at $z=0$ is shown in panel (a). Panels (b,c) show the distribution for halos located in regions with mean total densities ($\Delta=\bar{\rho}_{m}[<1\Mpc/h]$, in units of cosmic mean) larger than 3 and 10, respectively. Vertical lines indicate the location of the 25\%, 50\%, and 75\% quartiles of each distribution. }
\label{fig:etahist}
\end{figure*}

Due to the highly biased nature of jet launching massive galaxies,
the resulting distribution of the external magnetic field in the IGM is also highly inhomogeneous. 
This indicates that the gas accretion suppression effect will be environmentally dependent.
To assess this expected systematic trend, we show results in panels (a,b,c) of 
Fig.~\ref{fig:etahist} for all environments (a), regions with overdensity $\Delta>3$ (b), 
and $\Delta>10$ (c), respectively,
where $\Delta$ is defined to be the total density within 1\,Mpc$/h$ of the halo center
in units of the mean total density at redshift $z=0$. 
In panel (a) we see
the $50\%_{-25\%}^{+25\%}$ quartiles of $\eta$ for halos of mass ranges $[10^{11},10^{12})$, $[10^{12},10^{13})$ and $[10^{13},10^{14})\msun$, respectively,
to be
$1.00^{+0.01}_{-0.06}$, $0.94^{+0.06}_{-0.17}$,
$0.83^{+0.17}_{-0.05}$.
In panel (b) we see
the $50\%_{-25\%}^{+25\%}$ quartiles of $\eta$ for halos of mass ranges $[10^{11},10^{12})$, $[10^{12},10^{13})$ and $[10^{13},10^{14})\msun$, respectively,
to be
$0.98^{+0.02}_{-0.13}$, $0.94^{+0.06}_{-0.17}$, 
$0.83^{+0.17}_{-0.05}$
for halos in regions with $\Delta > 3$, 
while they become
$0.92^{+0.08}_{-0.33}$, 
$0.86^{+0.11}_{-0.11}$, 
$0.83^{+0.17}_{-0.05}$
for halos in regions with $\Delta > 10$ (panel c).
A tabulated $\eta$ distribution in various halo mass ranges and overdensities is given in Table~\ref{tab:eta}. The increased impact on baryon reduction indicates that the ``global'' feedback is expected to be more pronounced in overdense
regions than in under-dense regions. 
In void regions, due to lack of massive galaxies, this feedback mechanism
may be absent, which is consistent with observations that galaxies
are bluest in voids at present \citep{Rojas2004}.

In Fig.~\ref{fig:etascatter},
we show $\eta$ as a function of the overdensity within the virial radius $\Delta$ (panel b),
the magnetic field strength at the virial radius $B_{\rm vir}$ (panel c)
and the inverse root of the plasma beta parameter at the virial radius $\beta^{-1/2}$ (panel d) for each galaxy. In regions with $B_{\rm vir}>0.01\mu$G, the average of $\eta$ is respectively 0.83, 0.86, and 0.88, in the mass ranges $[10^{11},10^{12})$, $[10^{12},10^{13})$ and $[10^{13},10^{14})\msun$.
Cen (2024) \citep{Cen2024} suggested that the suppression of baryon fraction 
may become significant when the magnetic pressure force exceeds the gravitational force at the virial radius of a halo, or equivalently, 
\begin{equation}
x \equiv {v_A\over \sqrt{2}\,\sigma_v} \gtrsim 1
\label{eqn:v}
\end{equation}
where $v_A$ is the \alf speed of the gas at the virial radius, and $\sigma_v$ is the 1-d velocity dispersion of \yq{all matter in} the halo. 
\yq{Alternatively, this relation would be equivalent to the thermal-to-magnetic pressure ratio, i.e., the plasma $\beta$
parameter
\begin{equation}
\begin{split}
\beta^{-1/2}= \sqrt{\frac{P_{\rm mag}}{P_{\rm th}}} \gtrsim 1
\end{split}
\label{eqn:v}
\end{equation}
at the virial radius, if the gas sound speed is equal to the halo velocity dispersion.
%
%
Given the widespread use of $\beta$ involving magnetic energy in many applications, we will adopt $\beta^{-1/2}$ instead of $x$.}
The physics underpinning the gas accretion suppression
is seen to be borne out in panel (d) of Fig.~\ref{fig:etascatter}, where $\eta$ begins to decline when $\beta^{-1/2}\gtrsim 0.1$.
For the lowest halo mass bin examined $[10^{11},10^{12})\msun$, it is seen that
this approximately corresponds to a magnetic field at the virial radius
of about $B_{\rm vir}\gtrsim 0.01\,\mu{\rm G}$ (panel c). 
Least-square fits to the three halo mass bins, in the form of $e^{-\alpha \beta^{-1/2}}$, are shown
as three curves. 
We see that, at a given $x$, the gas suppression effects are stronger
for higher mass halos.
This is in a large part due to the fact that higher mass halos accrete fractionally
more mass than lower mass halos since the injection of magnetic bubbles in
the simulation at $z=2$. 
Because the magnetic pressure induced suppression effects are external in nature,
already accreted gas in halos is not much affected.
For example, some low mass halos have $B_{\rm vir}>0.1\mu$G, but with $\eta>0.9$. 
These halos either already have assembled most of their final masses before $z=2$, i.e., before radio lobes are implemented, or have cosmic filament orientation aligned with the magnetic field direction, which enables continued supply of baryons into the galaxy via filaments
without hindrance from the magnetic pressure force. 
Needless to say, the gas suppression effects due to this mechanism and associated effects
on star formation and others are complex. Detailed simulatons that include this 
feedback process and other traditional internal feedback processes
are likely to operate in a complex, intertwined, probably synergistic fashion.

In conclusion, in this article we use cosmological magneto-hydrodynamic simulations to quantify the impact giant radio lobes have on the accretion of gas onto galaxies in the greater neighborhood. The radio lobes are modeled as magnetic energy injection in the form of megaparsec-scale giant bubbles,
which are matched with observational constraints in both bubble size
and energetics, at cosmic noon ($z=2$). 
We show that, due to the highly biased nature of radio-lobe launching massive galaxies,
the effects on gas accretion onto galaxies are strongly environment dependent.
Even in the same environment, such as parameterized by the matter overdensity at certain large scale,
the effects are variable.
Furthermore, when gauged by the ratio of the 
\yq{plasma beta parameter at the virial radius}
of the halo, the effects on the overall gas suppression are also varied,
which may be due in part to the various assembly histories halos since the injection
of the radio bubbles at $z=2$ and in part to complex magnetic pressure dynamics
(such as orientation of the field to the orientation of the gas-feeding cosmic filaments, etc).
Nonetheless, we find an expected, strong trend that the gas suppression effect
increases with 
the inverse root of the plasma beta parameter at the virial radius
of the halo (Fig.~\ref{fig:etascatter}).
Our analysis demonstrates that magnetic fields seeded by relativistic jets are capable of significantly impacting the gas accretion of nearby galaxies.
It is worth noting that some feedback strengths for internal AGN feedback
implemented in current cosmological galaxy formation simulations 
may be too large to be concordant with the observed AGN feedback
\citep[e.g., outflow kinetic luminosity in][]{2022Molina}.
The external, radio-lobe generated AGN feedback, combined with a more realistic internal AGN feedback model, may improve the physical realism of the AGN feedback energetics in simulations, as well as possess the potential for better matching observations. 
\yq{Future observations that probe the magnetic universe, such as from SKA \citep{Heald2020}, will be able help elucidate the impact of magnetic fields on the accretion of baryons onto galaxies in two ways. First, a larger population of high-redshift sources with rotation measures (RM) will greatly help probe the magnetic field in proto-clusters. Second, the high sensitivities of SKA may be able to directly detect synchrotron radiation of radio lobes at high redshift,
as recent detection of synchrotron radiation of z=0.9 radio jets indicates \citep{Oei2024}.}
We argue for inclusion of this new, global feedback process in the next generation of cosmological simulations. 

\begin{table*}[th!]
\centering
\caption{Tabulated distribution of $\eta$, expressed as the fraction of halos ($f_h$) with $\eta$ below a threshold. For each halo mass range, different columns show the fractions for halos located in different environmental overdensities $\Delta$, defined as the mean total density within 1\,Mpc/h from each halo center, in units of the cosmic mean at $z=0$. }
\begin{tabular}{c|rrr|rrr|rrr}
\toprule
  & \multicolumn{3}{c|}{$M_h\in$\,[$10^{11}$, $10^{12}$)\,$M_\odot$} & \multicolumn{3}{c|}{$M_h\in$\,[$10^{12}$, $10^{13}$)\,$M_\odot$} & \multicolumn{3}{c}{$M_h\in$\,[$10^{13}$, $10^{14}$)\,$M_\odot$} \\  \cline{2-10} 
 & \multicolumn{1}{c|}{$\Delta>0$} & \multicolumn{1}{c|}{$\Delta>3$} & \multicolumn{1}{c|}{$\Delta>10$} & \multicolumn{1}{c|}{$\Delta>0$} & \multicolumn{1}{c|}{$\Delta>3$} & \multicolumn{1}{c|}{$\Delta>10$} & \multicolumn{1}{c|}{$\Delta>0$} & \multicolumn{1}{c|}{$\Delta>3$} & \multicolumn{1}{c}{$\Delta>10$} \\ \midrule
$f_h$ ($\eta<0.95$) & \multicolumn{1}{l|}{0.27} & \multicolumn{1}{l|}{0.39} & \multicolumn{1}{l|}{0.60} & \multicolumn{1}{l|}{0.50} & \multicolumn{1}{l|}{0.50} & \multicolumn{1}{l|}{0.71} & \multicolumn{1}{l|}{0.71} & \multicolumn{1}{l|}{0.71} & \multicolumn{1}{l}{0.71} \\
$f_h$ ($\eta<0.90$) & \multicolumn{1}{l|}{0.20} & \multicolumn{1}{l|}{0.27} & \multicolumn{1}{l|}{0.40} & \multicolumn{1}{l|}{0.42} & \multicolumn{1}{l|}{0.42} & \multicolumn{1}{l|}{0.57} & \multicolumn{1}{l|}{0.57} & \multicolumn{1}{l|}{0.57} & \multicolumn{1}{l}{0.57} \\
$f_h$ ($\eta<0.85$) & \multicolumn{1}{l|}{0.15} & \multicolumn{1}{l|}{0.21} & \multicolumn{1}{l|}{0.25} & \multicolumn{1}{l|}{0.35} & \multicolumn{1}{l|}{0.35} & \multicolumn{1}{l|}{0.43} & \multicolumn{1}{l|}{0.57} & \multicolumn{1}{l|}{0.57} & \multicolumn{1}{l}{0.57} \\
$f_h$ ($\eta<0.80$) & \multicolumn{1}{l|}{0.12} & \multicolumn{1}{l|}{0.17} & \multicolumn{1}{l|}{0.25} & \multicolumn{1}{l|}{0.27} & \multicolumn{1}{l|}{0.27} & \multicolumn{1}{l|}{0.43} & \multicolumn{1}{l|}{0.29} & \multicolumn{1}{l|}{0.29} & \multicolumn{1}{l}{0.29} \\
$f_h$ ($\eta<0.75$) & \multicolumn{1}{l|}{0.095} & \multicolumn{1}{l|}{0.14} & \multicolumn{1}{l|}{0.25} & \multicolumn{1}{l|}{0.15} & \multicolumn{1}{l|}{0.15} & \multicolumn{1}{l|}{0.21} & \multicolumn{1}{l|}{0} & \multicolumn{1}{l|}{0} & \multicolumn{1}{l}{0} \\
$f_h$ ($\eta<0.70$) & \multicolumn{1}{l|}{0.079} & \multicolumn{1}{l|}{0.11} & \multicolumn{1}{l|}{0.20} & \multicolumn{1}{l|}{0.038} & \multicolumn{1}{l|}{0.038} & \multicolumn{1}{l|}{0.071} & \multicolumn{1}{l|}{0} & \multicolumn{1}{l|}{0} & \multicolumn{1}{l}{0} \\
$f_h$ ($\eta<0.60$) & \multicolumn{1}{l|}{0.043} & \multicolumn{1}{l|}{0.063} & \multicolumn{1}{l|}{0.15} & \multicolumn{1}{l|}{0} & \multicolumn{1}{l|}{0} & \multicolumn{1}{l|}{0} & \multicolumn{1}{l|}{0} & \multicolumn{1}{l|}{0} & \multicolumn{1}{l}{0} \\
$f_h$ ($\eta<0.50$) & \multicolumn{1}{l|}{0.012} & \multicolumn{1}{l|}{0.014} & \multicolumn{1}{l|}{0} & \multicolumn{1}{l|}{0} & \multicolumn{1}{l|}{0} & \multicolumn{1}{l|}{0} & \multicolumn{1}{l|}{0} & \multicolumn{1}{l|}{0} & \multicolumn{1}{l}{0} \\ \bottomrule
\end{tabular}
\label{tab:eta}
\end{table*}

\acknow{
This work is supported by the National Key Research and Development Program of China and the Zhejiang provincial top level research support program. 
The simulations and analysis presented in this article were carried out on the SilkRiver Supercomputer of Zhejiang University.
We thank the High Performance Computing Platform at Zhejiang Lab, where part of the simulation modules were developed.
Analysis was performed using the yt\_astro\_analysis extension \citep{yt.astro.analysis} of the yt analysis toolkit \citep{yt}.
}

\showacknow{} 



\begin{thebibliography}{}

\end{thebibliography}


\begin{thebibliography}{10}

\bibitem{Grasso2001}
D {Grasso}, HR {Rubinstein}, {Magnetic fields in the early Universe}.
\newblock {\em\protect\JournalTitle{Physics Reports}} \textbf{348}, 163--266 (2001).

\bibitem{Govoni2004}
F {Govoni}, L {Feretti}, {Magnetic Fields in Clusters of Galaxies}.
\newblock {\em\protect\JournalTitle{International Journal of Modern Physics D}}
  \textbf{13}, 1549--1594 (2004).

\bibitem{Beck2015}
R {Beck}, {Magnetic fields in spiral galaxies}.
\newblock {\em\protect\JournalTitle{\aapr}} \textbf{24}, 4 (2015).

\bibitem{Feretti2012}
L {Feretti}, G {Giovannini}, F {Govoni}, M {Murgia}, {Clusters of galaxies:
  observational properties of the diffuse radio emission}.
\newblock {\em\protect\JournalTitle{\aapr}} \textbf{20}, 54 (2012).

\bibitem{Han2017}
JL {Han}, {Observing Interstellar and Intergalactic Magnetic Fields}.
\newblock {\em\protect\JournalTitle{\araa}} \textbf{55}, 111--157 (2017).

\bibitem{Ratra1992}
B {Ratra}, {Cosmological ``Seed'' Magnetic Field from Inflation}.
\newblock {\em\protect\JournalTitle{\apjl}} \textbf{391}, L1 (1992).

\bibitem{Durrer2013}
R {Durrer}, A {Neronov}, {Cosmological magnetic fields: their generation,
  evolution and observation}.
\newblock {\em\protect\JournalTitle{\aapr}} \textbf{21}, 62 (2013).

\bibitem{Subramanian2016}
K {Subramanian}, {The origin, evolution and signatures of primordial magnetic
  fields}.
\newblock {\em\protect\JournalTitle{Reports on Progress in Physics}}
  \textbf{79}, 076901 (2016).

\bibitem{1997Kulsrud}
RM {Kulsrud}, R {Cen}, JP {Ostriker}, D {Ryu}, {The Protogalactic Origin for
  Cosmic Magnetic Fields}.
\newblock {\em\protect\JournalTitle{\apj}} \textbf{480}, 481 (1997).

\bibitem{Brandenburg2005}
A {Brandenburg}, K {Subramanian}, {Astrophysical magnetic fields and nonlinear
  dynamo theory}.
\newblock {\em\protect\JournalTitle{Physics Reports}} \textbf{417}, 1--209 (2005).

\bibitem{Vazza2014}
F {Vazza}, M {Br{\"u}ggen}, C {Gheller}, P {Wang}, {On the amplification of
  magnetic fields in cosmic filaments and galaxy clusters}.
\newblock {\em\protect\JournalTitle{\mnras}} \textbf{445}, 3706--3722 (2014).

\bibitem{2018Springel}
V {Springel}, et~al., {First results from the IllustrisTNG simulations: matter
  and galaxy clustering}.
\newblock {\em\protect\JournalTitle{\mnras}} \textbf{475}, 676--698 (2018).

\bibitem{Katz2021}
H {Katz}, et~al., {Introducing SPHINX-MHD: the impact of primordial magnetic
  fields on the first galaxies, reionization, and the global 21-cm signal}.
\newblock {\em\protect\JournalTitle{\mnras}} \textbf{507}, 1254--1282 (2021).

\bibitem{Bertone2006}
S {Bertone}, C {Vogt}, T {En{\ss}lin}, {Magnetic field seeding by galactic
  winds}.
\newblock {\em\protect\JournalTitle{\mnras}} \textbf{370}, 319--330 (2006).

\bibitem{Donnert2009}
J {Donnert}, K {Dolag}, H {Lesch}, E {M{\"u}ller}, {Cluster magnetic fields
  from galactic outflows}.
\newblock {\em\protect\JournalTitle{\mnras}} \textbf{392}, 1008--1021 (2009).

\bibitem{Cen2024}
R {Cen}, {Global preventive feedback of powerful radio jets on galaxy
  formation}.
\newblock {\em\protect\JournalTitle{Proceedings of the National Academy of
  Science}} \textbf{121}, e2402435121 (2024).

\bibitem{Ruszkowski2023}
M {Ruszkowski}, C {Pfrommer}, {Cosmic ray feedback in galaxies and galaxy
  clusters}.
\newblock {\em\protect\JournalTitle{\aapr}} \textbf{31}, 4 (2023).

\bibitem{Quataert2025}
E {Quataert}, PF {Hopkins}, {Cosmic Ray Feedback in Massive Halos: Implications
  for the Distribution of Baryons}.
\newblock {\em\protect\JournalTitle{The Open Journal of Astrophysics}}
  \textbf{8}, 66 (2025).

\bibitem{2014Vogelsberger}
M {Vogelsberger}, et~al., {Introducing the Illustris Project: simulating the
  coevolution of dark and visible matter in the Universe}.
\newblock {\em\protect\JournalTitle{\mnras}} \textbf{444}, 1518--1547 (2014).

\bibitem{2014Genel}
S {Genel}, et~al., {Introducing the Illustris project: the evolution of galaxy
  populations across cosmic time}.
\newblock {\em\protect\JournalTitle{\mnras}} \textbf{445}, 175--200 (2014).

\bibitem{2015Schaye}
J {Schaye}, et~al., {The EAGLE project: simulating the evolution and assembly
  of galaxies and their environments}.
\newblock {\em\protect\JournalTitle{\mnras}} \textbf{446}, 521--554 (2015).

\bibitem{2015Crain}
RA {Crain}, et~al., {The EAGLE simulations of galaxy formation: calibration of
  subgrid physics and model variations}.
\newblock {\em\protect\JournalTitle{\mnras}} \textbf{450}, 1937--1961 (2015).

\bibitem{2015Schaller}
M {Schaller}, et~al., {Baryon effects on the internal structure of
  {\ensuremath{\Lambda}}CDM haloes in the EAGLE simulations}.
\newblock {\em\protect\JournalTitle{\mnras}} \textbf{451}, 1247--1267 (2015).

\bibitem{2015Trayford}
JW {Trayford}, et~al., {Colours and luminosities of z = 0.1 galaxies in the
  EAGLE simulation}.
\newblock {\em\protect\JournalTitle{\mnras}} \textbf{452}, 2879--2896 (2015).

\bibitem{2016McAlpine}
S {McAlpine}, et~al., {The EAGLE simulations of galaxy formation: Public
  release of halo and galaxy catalogues}.
\newblock {\em\protect\JournalTitle{Astronomy and Computing}} \textbf{15},
  72--89 (2016).

\bibitem{2018Pillepich}
A {Pillepich}, et~al., {Simulating galaxy formation with the IllustrisTNG
  model}.
\newblock {\em\protect\JournalTitle{\mnras}} \textbf{473}, 4077--4106 (2018).

\bibitem{2018Nelson}
D {Nelson}, et~al., {First results from the IllustrisTNG simulations: the
  galaxy colour bimodality}.
\newblock {\em\protect\JournalTitle{\mnras}} \textbf{475}, 624--647 (2018).

\bibitem{2019Dave}
R {Dav{\'e}}, et~al., {SIMBA: Cosmological simulations with black hole growth
  and feedback}.
\newblock {\em\protect\JournalTitle{\mnras}} \textbf{486}, 2827--2849 (2019).

\bibitem{2022Sorini}
D {Sorini}, R {Dav{\'e}}, W {Cui}, S {Appleby}, {How baryons affect haloes and
  large-scale structure: a unified picture from the SIMBA simulation}.
\newblock {\em\protect\JournalTitle{\mnras}} \textbf{516}, 883--906 (2022).

\bibitem{Bryan2014}
GL {Bryan}, et~al., {ENZO: An Adaptive Mesh Refinement Code for Astrophysics}.
\newblock {\em\protect\JournalTitle{\apjs}} \textbf{211}, 19 (2014).

\bibitem{Li2008}
S {Li}, H {Li}, R {Cen}, {CosmoMHD: A Cosmological Magnetohydrodynamics Code}.
\newblock {\em\protect\JournalTitle{\apjs}} \textbf{174}, 1--12 (2008).

\bibitem{Collins2010}
DC {Collins}, H {Xu}, ML {Norman}, H {Li}, S {Li}, {Cosmological Adaptive Mesh
  Refinement Magnetohydrodynamics with Enzo}.
\newblock {\em\protect\JournalTitle{\apjs}} \textbf{186}, 308--333 (2010).

\bibitem{Hahn2011}
O {Hahn}, T {Abel}, {Multi-scale initial conditions for cosmological
  simulations}.
\newblock {\em\protect\JournalTitle{\mnras}} \textbf{415}, 2101--2121 (2011).

\bibitem{Aghanim2020}
{Planck Collaboration}, et~al., {Planck 2018 results. VI. Cosmological
  parameters}.
\newblock {\em\protect\JournalTitle{\aap}} \textbf{641}, A6 (2020).

\bibitem{Smith2017a}
BD {Smith}, et~al., {GRACKLE: a chemistry and cooling library for
  astrophysics}.
\newblock {\em\protect\JournalTitle{\mnras}} \textbf{466}, 2217--2234 (2017).

\bibitem{Behroozi2013}
PS {Behroozi}, RH {Wechsler}, HY {Wu}, {The ROCKSTAR Phase-space Temporal Halo
  Finder and the Velocity Offsets of Cluster Cores}.
\newblock {\em\protect\JournalTitle{\apj}} \textbf{762}, 109 (2013).

\bibitem{ODea2009}
CP {O'Dea}, RA {Daly}, P {Kharb}, KA {Freeman}, SA {Baum}, {Physical properties
  of very powerful FRII radio galaxies}.
\newblock {\em\protect\JournalTitle{\aap}} \textbf{494}, 471--488 (2009).

\bibitem{Miley1980}
G {Miley}, {The structure of extended extragalactic radio sources}.
\newblock {\em\protect\JournalTitle{\araa}} \textbf{18}, 165--218 (1980).

\bibitem{Hardcastle2000}
MJ {Hardcastle}, DM {Worrall}, {The environments of FRII radio sources}.
\newblock {\em\protect\JournalTitle{\mnras}} \textbf{319}, 562--572 (2000).

\bibitem{Tchekhovskoy2016}
A {Tchekhovskoy}, O {Bromberg}, {Three-dimensional relativistic MHD simulations
  of active galactic nuclei jets: magnetic kink instability and Fanaroff-Riley
  dichotomy}.
\newblock {\em\protect\JournalTitle{\mnras}} \textbf{461}, L46--L50 (2016).

\bibitem{Vazza2021}
F {Vazza}, D {Wittor}, G {Brunetti}, M {Br{\"u}ggen}, {Simulating the transport
  of relativistic electrons and magnetic fields injected by radio galaxies in
  the intracluster medium}.
\newblock {\em\protect\JournalTitle{\aap}} \textbf{653}, A23 (2021).

\bibitem{Kravtsov2005}
AV {Kravtsov}, D {Nagai}, AA {Vikhlinin}, {Effects of Cooling and Star
  Formation on the Baryon Fractions in Clusters}.
\newblock {\em\protect\JournalTitle{\apj}} \textbf{625}, 588--598 (2005).

\bibitem{Rojas2004}
RR {Rojas}, MS {Vogeley}, F {Hoyle}, J {Brinkmann}, {Photometric Properties of
  Void Galaxies in the Sloan Digital Sky Survey}.
\newblock {\em\protect\JournalTitle{\apj}} \textbf{617}, 50--63 (2004).

\bibitem{2022Molina}
J {Molina}, et~al., {Ionized Outflows in Nearby Quasars Are Poorly Coupled to
  Their Host Galaxies}.
\newblock {\em\protect\JournalTitle{\apj}} \textbf{935}, 72 (2022).

\bibitem{Heald2020}
G {Heald}, et~al., {Magnetism Science with the Square Kilometre Array}.
\newblock {\em\protect\JournalTitle{Galaxies}} \textbf{8}, 53 (2020).

\bibitem{Oei2024}
MSSL {Oei}, et~al., {Black hole jets on the scale of the cosmic web}.
\newblock {\em\protect\JournalTitle{\nat}} \textbf{633}, 537--541 (2024).

\bibitem{yt.astro.analysis}
B {Smith}, et~al., {yt-project/yt\_astro\_analysis: yt\_astro\_analysis version
  1.1.1 is released!} (2022).

\bibitem{yt}
MJ {Turk}, et~al., {yt: A Multi-code Analysis Toolkit for Astrophysical
  Simulation Data}.
\newblock {\em\protect\JournalTitle{\apjs}} \textbf{192}, 9 (2011).

\end{thebibliography}

\end{document}